\documentclass[sigconf,screen]{acmart}

\usepackage[ruled,vlined,lined,commentsnumbered]{algorithm2e}
\usepackage{amsmath,amsfonts}
\usepackage{array}
\usepackage{booktabs}
\usepackage[skip=1pt,labelfont=bf]{caption}
 \usepackage{calligra}
\usepackage{color, colortbl}
\usepackage{courier}
\usepackage{csvsimple}
\usepackage{enumitem}
\usepackage{fancybox}
\usepackage{fontenc}
\usepackage{graphicx}
\usepackage{listings}
\usepackage{longtable}
\usepackage{lscape}
\usepackage{makecell}
\usepackage{marvosym}
\usepackage{moreverb}
\usepackage{multicol}
\usepackage{multirow}
\usepackage{pifont}
\usepackage{rotating}
\usepackage{setspace}
\usepackage{caption}
\usepackage{subfigure}
\usepackage[most]{tcolorbox}
\usepackage{threeparttable}
\usepackage{tikz}
\usepackage[normalem]{ulem}
\usepackage{url}
\usepackage{soul}
\usepackage{wasysym}
\usepackage{svg}
\usepackage{textcomp}
\usepackage{xcolor}
\usepackage{wrapfig}
\usepackage{pdflscape}
\usepackage{hyperref}

\usepackage{amsmath}
\usepackage{array}
\usepackage{balance}
\usepackage{microtype}

\newcolumntype{L}[1]{>{\raggedright\let\newline\\\arraybackslash\hspace{0pt}}m{#1}}
\newcolumntype{C}[1]{>{\centering\let\newline\\\arraybackslash\hspace{0pt}}m{#1}}
\newcolumntype{R}[1]{>{\raggedleft\let\newline\\\arraybackslash\hspace{0pt}}m{#1}}

\newboolean{showcomments}
\setboolean{showcomments}{true}
\ifthenelse{\boolean{showcomments}}
 { \newcommand{\mynote}[2]{
      \fbox{\bfseries\sffamily\scriptsize#1}
        {\small$\blacktriangleright$\textsf{\emph{#2}}$\blacktriangleleft$}}}
        { \newcommand{\mynote}[2]{}}

\newcommand{\toolname}{SVulD\xspace}
\newcommand{\toolnameB}{\textbf{SVulD}}

\definecolor{lightgray}{gray}{0.9}

\newcommand{\intuition}[1]{
\begin{tcolorbox}[colback=white,boxrule=1pt,top=0pt,bottom=0pt,left=1pt,right=2pt,top=2pt,bottom=2pt]
\em #1
\end{tcolorbox}
}

\AtBeginDocument{%
  \providecommand\BibTeX{{%
    \normalfont B\kern-0.5em{\scshape i\kern-0.25em b}\kern-0.8em\TeX}}}

\setcopyright{acmlicensed}
\acmPrice{15.00}
\acmDOI{10.1145/3611643.3616358}
\acmYear{2023}
\copyrightyear{2023}
\acmSubmissionID{fse23main-p1405-p}
\acmISBN{979-8-4007-0327-0/23/12}
\acmConference[ESEC/FSE '23]{Proceedings of the 31st ACM Joint European Software Engineering Conference and Symposium on the Foundations of Software Engineering}{December 3--9, 2023}{San Francisco, CA, USA}
\acmBooktitle{Proceedings of the 31st ACM Joint European Software Engineering Conference and Symposium on the Foundations of Software Engineering (ESEC/FSE '23), December 3--9, 2023, San Francisco, CA, USA}
\received{2023-03-02}
\received[accepted]{2023-07-27}

\begin{document}

\title[Distinguishing Look-Alike Innocent and Vulnerable Code...]{
Distinguishing Look-Alike Innocent and Vulnerable Code by Subtle Semantic Representation Learning and Explanation
}

\author{Chao Ni}
\affiliation{%
  \institution{School of Software Technology, Zhejiang University}
  \city{Hangzhou}
 \state{Zhejiang}
  \country{China}
}
\email{chaoni@zju.edu.cn}

\author{Xin Yin}
\affiliation{%
  \institution{School of Software Technology, Zhejiang University}
  \city{Hangzhou}
  \state{Zhejiang}
  \country{China}}
  \email{xyin@zju.edu.cn}

\author{Kaiwen Yang}
\affiliation{%
  \institution{College of  Computer Science and Technology, Zhejiang University}
  \city{Hangzhou}
  \state{Zhejiang}
  \country{China}}
\email{kwyang@zju.edu.cn}

\author{Dehai Zhao}
\affiliation{%
  \institution{Data61, CSIRO}
  \city{Sydney}
  \country{Australia}
  }
\email{dehai.zhao@data61.csiro.au}

\author{Zhenchang Xing}
\affiliation{%
  \institution{Research School of Computer Science, Australian National University and Data61, CSIRO}
  \city{Canberra}
  \country{Australia}
  }
\email{zhenchang.xing@anu.edu.au}
 
\author{Xin Xia}
\authornote{Xin Xia is the corresponding author.}
\affiliation{%
   \institution{Zhejiang University}
  \city{Hangzhou}
  \state{Zhejiang}
  \country{China}
  }
\email{xin.xia@acm.org}

\begin{abstract}


Though many deep learning (DL)-based vulnerability detection approaches have been proposed and indeed achieved remarkable performance, they still have limitations in the generalization as well as the practical usage.
More precisely, existing DL-based approaches (1) perform negatively on prediction tasks among functions that are lexically similar but have contrary semantics; (2) provide no intuitive developer-oriented explanations to the detected results.

In this paper, we propose a novel approach named \toolname, a function-level \underline{S}ubtle semantic embedding for \underline{Vul}nerability \underline{D}etection along with intuitive explanations, to alleviate the above limitations.
Specifically, \toolname firstly trains a model to learn distinguishing semantic representations of functions regardless of their lexical similarity.
Then, for the detected vulnerable functions, \toolname provides natural language explanations (e.g., \textit{root cause}) of results to help developers intuitively understand the vulnerabilities.
To evaluate the effectiveness of \toolname, we conduct large-scale experiments on a widely used practical vulnerability dataset and compare it with four state-of-the-art (SOTA) approaches by considering five performance measures.
The experimental results indicate that \toolname outperforms all SOTAs with a substantial improvement (i.e., 23.5\%-68.0\% in terms of F1-score,  15.9\%-134.8\% in terms of PR-AUC and 7.4\%-64.4\% in terms of Accuracy).
Besides, we conduct a user-case study to evaluate the usefulness of \toolname for developers on understanding the vulnerable code and the participants' feedback demonstrates that \toolname is helpful for development practice.

\end{abstract}



\maketitle

\section{Introduction}

Software vulnerabilities have caused massive damage to software systems and many automatic vulnerability detection approaches have been proposed to prevent software systems from severity attacks and indeed achieved promising results, which can be broadly classified into two categories: static analysis approaches~\cite{rats,Checkmarx,FlawFinder,Coverity,fan2019smoke,li2020pca} and deep learning (DL) approaches~\cite{cao2022mvd,li2018vuldeepecker,zhou2019devign,cheng2021deepwukong,li2021sysevr,yamaguchi2014modeling,li2021vuldeelocator,duan2019vulsniper,lin2017poster,chakraborty2021deep,wu2022vulcnn}.
The static analysis approaches focus on detecting type-specific vulnerabilities (i.e., user-after-free) with the help of user-defined rules or patterns, which highly depend on expert knowledge and have little chance to find a wider range of vulnerabilities~\cite{cao2022mvd,cheng2022path}.
The deep learning approaches, benefiting from the powerful learning ability of deep neural networks, aim at leveraging advanced models to capture program semantics to identify potential type-agnostic software vulnerabilities. 
That is, these approaches automatically extract implicit vulnerability patterns from previous vulnerable code instead of requiring expert involvement, 
which makes deep learning become a good choice to solve vulnerability detection problems.
However, the existing DL-based approaches still have two limitations that affect their effectiveness of generalization and the usefulness of development practice.

The first problem is that existing DL-based approaches have limited ability to distinguish subtle semantic differences among lexically similar functions.
For a specific version of a vulnerable function, the vulnerabilities are usually fixed with a few modifications to it (i.e., 52.6\% vulnerable functions can be fixed within 5 (76.7\% within 10) lines of code in our dataset).
The fixed functions can be conceptually treated as \textit{non-vulnerable} functions.
Meanwhile, we find that \textit{the vulnerable function and its corresponding fixed function are extremely lexically similar (i.e., fixing a vulnerability by modifying less than 100 CHARs accounts for 46.0\% (200 for 65.1\%)) but they have significant semantic differences (i.e., vulnerable or non-vulnerable)}.
Ideally,  we expect that a good-performing DL-based approach can perform equally well in detecting vulnerable functions and their corresponding fixing patches.
However, we find that the SOTA DL-based approaches perform negatively on the fixed functions (i.e., non-vulnerable ones) and incorrectly classify the fixed version as vulnerable ones (43.5\%-63.1\% false positive).
Thus, it is urgently required to pay more attention to semantic differences among lexically similar functions with contrasting semantics.

The second problem is that existing vulnerability detection approaches focus on giving binary detection results (i.e., vulnerable or not) and ignore the importance of providing developer-oriented natural-language explanations for the results.
For example, \textit{what is the possible root cause of such vulnerability?} \textit{what impacts will be caused by this vulnerability?}
Those explanations may help developers to have a better understanding of the detected vulnerable code.
However, considering the concealment of software vulnerabilities, it is hard to observe two identical vulnerabilities.
It is believed that similar/homogeneous vulnerabilities have similar root causes or lead to similar impacts.
Intuitively, we find that many publicly available developer forums (i.e., Stack Overflow) share semantically similar problematic source code, and some of the responses provide useful and understandable natural language explanations about the issues, which help developers to intuitively figure out the potential root cause inside their problematic code.

To mitigate the above two limitations, we propose a novel approach named \toolname, which is a function-level \underline{S}ubtle semantic embedding for \underline{Vul}nerability \underline{D}etection along with intuitive explanations.
It is technically based on pre-trained semantic embedding~\cite{guo2022unixcoder} as well as contrastive learning~\cite{chen2020simple}.
Specifically, to solve the first issue, \toolname adopts contrastive learning to train the UniXcoder~\cite{guo2022unixcoder} semantic embedding model in order to learn the semantic representation of functions regardless of their lexically similar information.
To address the second issue, we build a knowledge-based crowdsource dataset by crawling problematic codes from Stack Overflow and fine-tune a BERT question-answering model~\cite{devlin2018bert,sun2021generating} on 1,678  manually labeled posts to automatically extract the key information from high-quality answers, which can provide developers with intuitive explanations and help them to understand the detected vulnerable  code.

To evaluate the effectiveness of \toolname, we conduct extensive experiments on widely used practical vulnerability dataset~\cite{li2021vulnerability,cheng2022path,nikitopoulos2021crossvul}.
Particularly, our \toolname is compared with four SOTA approaches (i.e., Devign, {\sc ReVeal}, {\sc IVDetect}, and LineVul) by five performance measures (i.e., Accuracy, Precision, Recall, F1-score, and PR-AUC).
The experimental results indicate that \toolname outperforms all SOTA baselines with a substantial improvement (i.e., 23.5\%-68.0\% in terms of F1-score,  15.9\%-134.8\% in terms of PR-AUC and 7.4\%-64.4\% in terms of Accuracy).
Besides, to provide developers with an intuitive explanation of the detected vulnerable code, we design a {quality-first sorting strategy} to prioritize the retrieved semantic-related post answers. 
We conduct a user-case study to evaluate whether our tool can help developers understand the problems in code intuitively and the participants' feedback demonstrates the usefulness of \toolname. 
Finally, this paper makes the main contributions as below:
\begin{itemize}[leftmargin=*]
    \item We propose \toolname, a novel function-level approach for vulnerability detection with intuitive explanations based on the pre-trained semantic embedding model, which leverages contrastive learning technology to obtain the distinguishing semantic representations among lexically similar functions.

    \item We comprehensively investigate the effectiveness of \toolname on vulnerability detection and the generalization of fixed functions.
    The experiment results indicate that \toolname outperforms SOTAs with a substantial improvement (e.g., 23.5\%-68.0\% in terms of F1-score, 15.9\%-134.8\% in terms of PR-AUC).
    Especially, \toolname has better generalization performance on fixed functions (e.g., 7.4\%-64.4\% in terms of Accuracy).

    \item To the best of our knowledge, we are first to provide an intuitive explanation of the results given by a vulnerability detection approach, and a user-case study  confirms the feasibility of intuitively explaining the results with crowdsourced knowledge.
\end{itemize}


\section{Motivating Example}
\label{sec:motivation}

\vspace{2mm}
\begin{figure}[t]
    \centerline{
        \includegraphics[width=\linewidth]{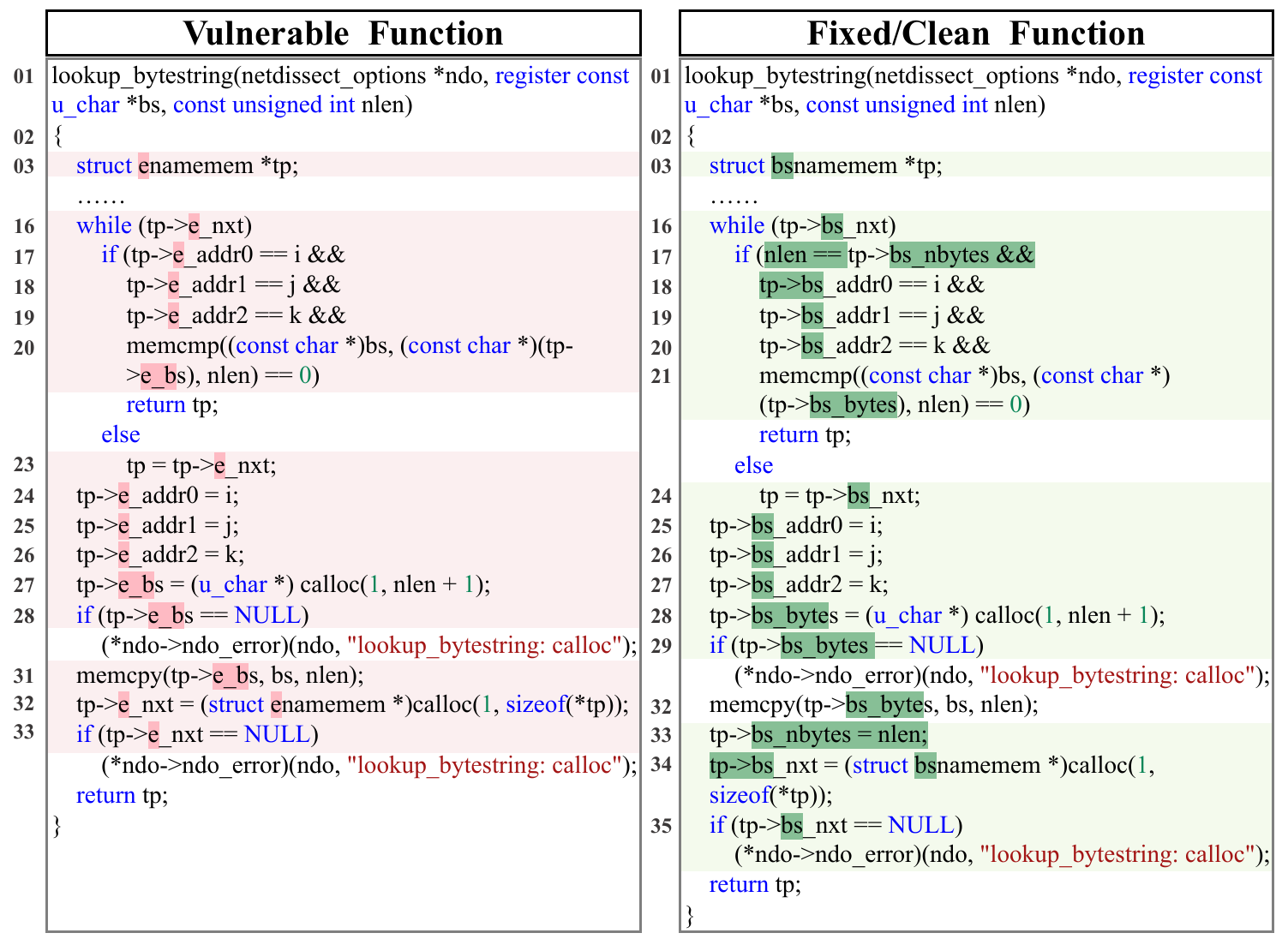}
    }
    \caption{
       An Out-of-bounds Read Vulnerability (CVE-2017-12894) in tcpdump}
    \label{fig:codeexample}
\end{figure}

Functions usually consist of several lines of code for implementing a specific program  semantic (i.e., functionality) and we use different labels (i.e., \textit{vulnerable}, \textit{non-vulnerable}) to describe the security status of functions.
A vulnerable function includes security defects (e.g., CWE-125: Out-of-bounds Read) in its codes, while a non-vulnerable function is clean.
A fixed function previously contains vulnerable codes but these codes have been fixed with some modifications on the vulnerable code snippets.
Therefore, the fixed functions can be conceptually treated as \textit{non-vulnerable} functions.

Fig.~\ref{fig:codeexample} shows two versions (the left one is for the vulnerable version, while the right one is for the non-vulnerable version) of a specific function in \textit{tcpdump} project~\cite{tcpdump}.
This function contains a typical \textit{Out-of-bounds Read} vulnerability CVE-2017-12894.
The \textit{if condition statement} does not detect the length of the address at line 17.
{\em Comparing the left vulnerable one with the right non-vulnerable/fixed one,
we find that the two versions are lexically similar but have distinguishing semantic differences from the security perspective}, which is not an accidental phenomenon.
We conduct statistical analysis about the vulnerable functions as well as their corresponding fixed functions on the widely used dataset named Big-Vul (10,900 vulnerable functions) collected by Fan et al.~\cite{fan2020ac} and find that 52.6\% vulnerable functions can be fixed within five lines of codes (LOCs, added or deleted lines) and 76.7\% functions can be fixed with less than 10 LOCs. 
From the view of modified chars, fixing a vulnerability by modifying less than 100 chars accounts for 46.0\% (200 chars for 65.1\%). 
Meanwhile, a function has a ratio of no more than 5\% accounts for  48.7\% (10\% for 63.4\%) between the number of modified chars and the whole number of chars.
All these statistical results indicate that the vulnerable function and the corresponding fixed function are extremely lexically similar.

Recently, benefiting from the powerful learning ability of deep neural networks, many SOTA DL-based vulnerability detection approaches (e.g., Devign~\cite{zhou2019devign}, {\sc ReVeal}~\cite{chakraborty2021deep}, {\sc IVDetect}~\cite{li2021vulnerability}, and LineVul~\cite{fu2022linevul}) have been proposed to capture program semantics in order to identify potential software vulnerabilities, and these approaches have achieved promising performance.
Ideally, a good-performing DL-based approach is expected to have a good generalization ability, which means that the approach should work well on both vulnerable and corresponding fixed non-vulnerable functions.
However, a large-scale experiment on Big-Vul shows that all these SOTA approaches have negative performance on predicting the fixed functions (i.e., non-vulnerable ones).
Specifically, they incorrectly classify the fixed functions as vulnerable ones (43.5\%-63.1\% false positive, cf. Section~\ref{lab:rq1} for details).

Meanwhile, almost all existing vulnerability detection approaches focus on classifying whether a function is vulnerable but do not provide developer-oriented natural-language explanations to help developers understand the detected vulnerable code.
For example, \textit{what is the possible root cause of such vulnerability?}
\textit{what impacts will be caused by this vulnerability?}
Such types of explanations may (at least intuitively) help developers to have a deeper understanding of the detected vulnerable code.
Intuitively, many publicly available user forums (i.e., Stack Overflow) share similar problematic source code and their corresponding responses may provide useful and understandable natural language explanations about the issues, which can intuitively help developers to figure out the potential root cause inside the vulnerable code.

\begin{figure}[!htbp]
    \centerline{
        \includegraphics[width=\linewidth]{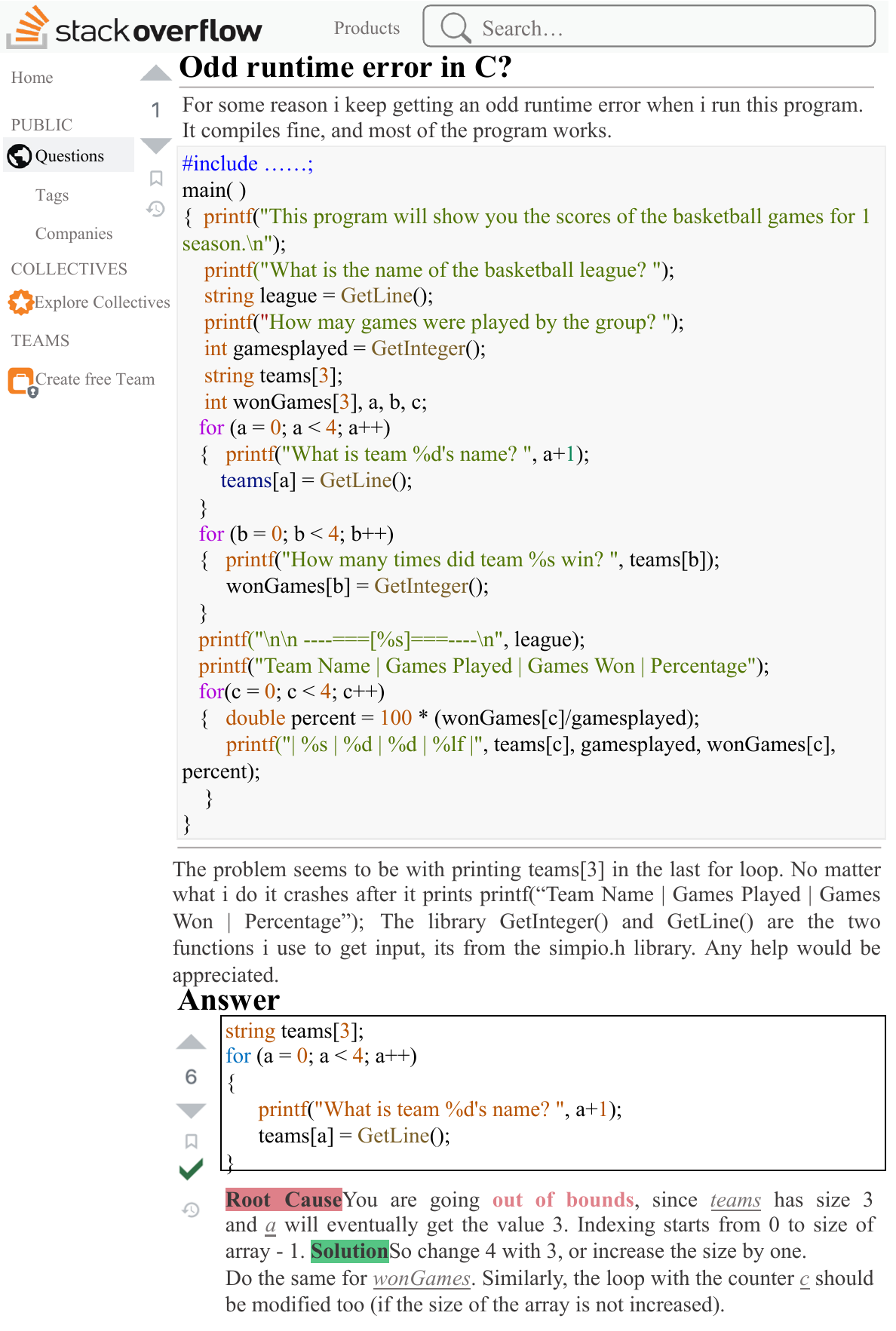}
    }
    \caption{
        A simple but similar problematic code along with an accepted answer in Stack Overflow. }
    \label{fig:soexample}
\end{figure}

As shown in Fig.~\ref{fig:soexample}, this code snippet has a similar root cause with the vulnerable function in Fig.~\ref{fig:codeexample}.
It crashes because of the limited size of defined arrays (i.e.,  \textit{teams} and \textit{wonGames}), which results in an \textit{Out-of-bounds} error when reading and writing content to the last element.
Similarly, the function in Fig.~\ref{fig:codeexample} will crash when the last element in their address array does not satisfy the length of a legal internet address.
If developers are provided with a natural language explanation of the root cause referring to the answer in Fig.~\ref{fig:soexample}, the problem in Fig.~\ref{fig:codeexample} will be easier to solve.

\intuition{\textbf{Motivating.} 
Two code snippets may be lexically similar but have distinct security semantics (vulnerable or non-vulnerable), which needs to embed their semantic difference in a better way.
Meanwhile, similar vulnerabilities may have a similar root cause, which can help participants understand the problematic codes better.
}

\vspace{2mm}

\section{Our Approach: \toolnameB
}
\label{sec:approach}

\begin{figure*}[!htbp]
\centerline{
    \includegraphics[width=.9\linewidth]{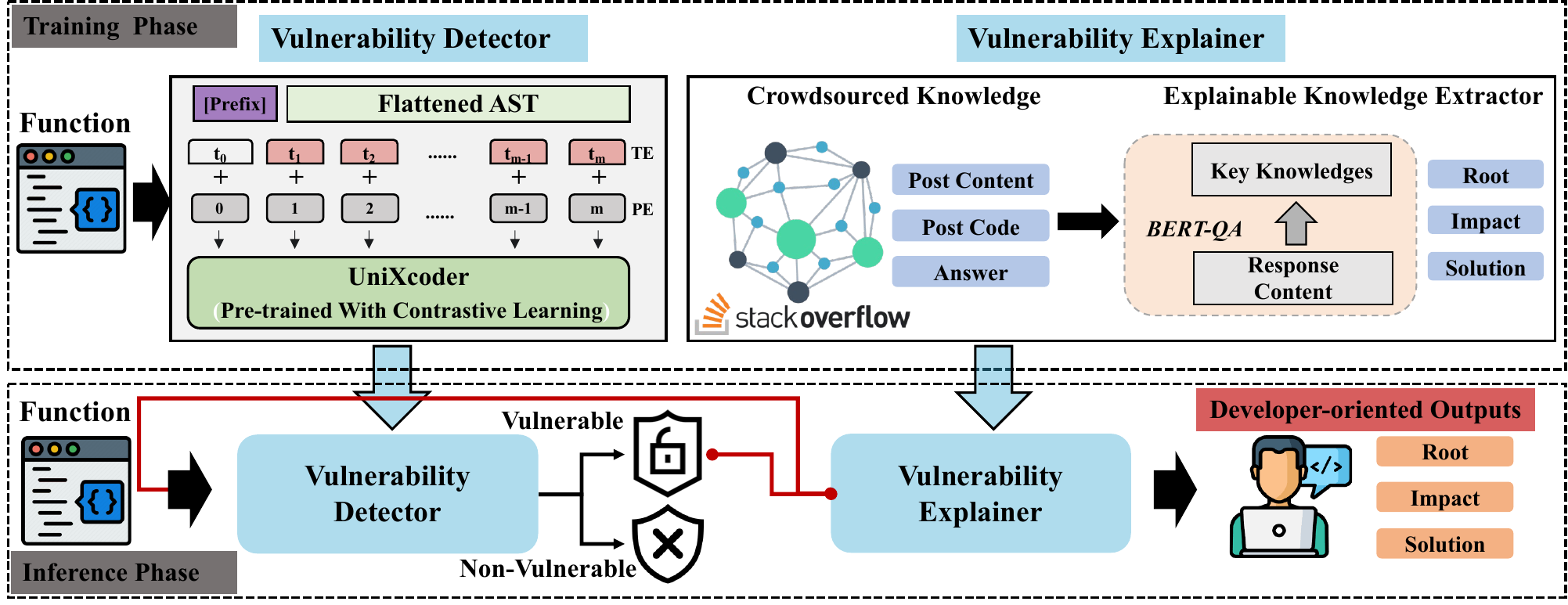}
    }
    \caption{The framework of \toolname.}
    \label{fig:framework}

\end{figure*}

To investigate the feasibility of our intuitive hypothesis, we propose a novel framework named \toolname, which integrates software vulnerability detection and intuitive natural language explanation.
As illustrated in Fig.~\ref{fig:framework}, \toolname consists of two main phases: \ding{182} training phase, where the vulnerability detector is trained on the high-quality dataset and vulnerability explainer is constructed on crowdsourced knowledge; \ding{183} inference phase, where a specific function is classified as vulnerable or not by the trained vulnerability detector and provide several developer-oriented explanations to the detected vulnerable function.
We present the details of \toolname in the following subsections.

\vspace{2mm}

\subsection{Vulnerability Detection}

In order to discriminate the semantic difference among lexically similar functions effectively, \toolname adopts contrastive learning framework with the pre-trained model, UniXcoder~\cite{guo2022unixcoder}, as the semantic encoder.
The architecture for contrastively training the UniXcoder-based semantic embedding model is illustrated in Fig.~\ref{fig:cl-detector}.
Contrastive learning~\cite{oord2018representation} is a kind of deep neural network training process that takes paired functions as input and uses the similarity between the paired functions as labels.
The training objective of contrastive learning is to learn whether two functions are
semantically similar regardless of their lexical similarity. 
Elaborately, the contrastive learning framework utilizes the encoder to embed source code into their semantic representations (i.e., hidden vectors) and aims at minimizing the distance between similar functions while maximizing the distance between dissimilar functions.
There are two important components of the proposed model: an encoder for embedding functions' semantics and a learning strategy for discriminating differences.

\subsubsection{Semantic Encoder}
Considering many successful applications of pre-trained models in software engineering (e.g., defect prediction~\cite{ni2022best} and code summarization~\cite{zhu2019automatic}), especially the recent work on vulnerability detection~\cite{fu2022linevul}, we leverage UniXcoder~\cite{guo2022unixcoder} as our semantic encoder.
It is a unified cross-modal (i.e., code, comment and abstract syntax tree (AST)) pre-trained model for programming language and utilizes mask attention matrices with prefix adapters (i.e., $[\mathit{prefix}]$) to control the behavior of the model (i.e., encoder only ($[\mathit{Enc}]$), decoder only ($[\mathit{Dec}]$) or encoder-decoder ($[\mathit{E2D}]$) ).
For each input function, UniXcoder encodes the AST of it into a sequence while retaining all structural information of the tree.
Meanwhile, in our binary classification setting, we set $[\mathit{prefix}]$ as $[\mathit{Enc}]$ and fine-tune it on our studied datasets to learn a better representation of source codes' semantic information.

\subsubsection{Semantic Difference Learning}
Our goal is to discriminate the semantic difference among lexical similar functions, which is consistent with the target of contrastive learning.
That is, minimize the distance between similar objects (i.e., the function in our study) while maximizing the distance between dissimilar objects.
Hoffer et al.~\cite{hoffer2015deep} proposed the triplet network for contrastive learning, which requires a triplet $(F, P, N)$ as the input, where $F$ corresponds to the
original source code of the function, $P$ refers to the positive equivalent of $F$, and $N$ is the negative one.
In our work, for a given function $F$ in the training data, its positive functions are the varying representation of the same functions and the negative functions are functions that are different from the given one.
Therefore, with a good semantic presentation, similar functions stay close to each other while dissimilar ones are far apart.

\begin{figure}[!h]
    \centerline{
        \includegraphics[width=\linewidth]{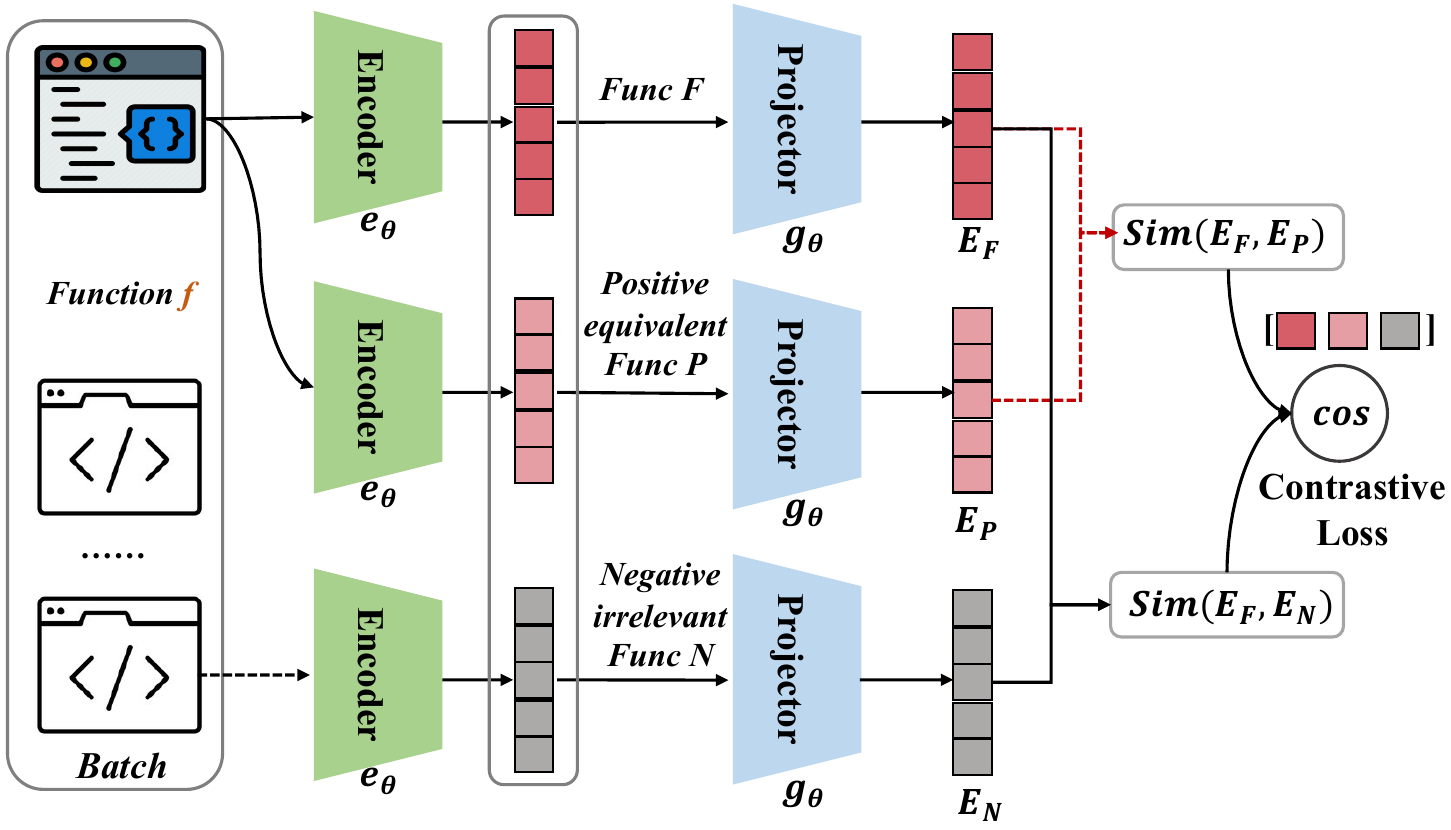}
    }
    \caption{Architecture for contrastively training UniXcoder
based semantic embedding model}
    \label{fig:cl-detector}
\end{figure}

Fig.~\ref{fig:cl-detector} shows the architecture of the contrastive learning used in this work, in which the UniXcoder is the base model for semantic embedding.
We use a Pooling layer to connect the UniXcoder model and the triple network. 
The triple network has two layers.
The first layer is three identical deep neural networks for feature extraction of input functions, which can be easily replaced with other semantic learning models.
The second layer of the triplet network is a loss function based on the cosine distance operator with transformation operations of \textit{projector}, which is used to minimize the distance between similar functions and maximize the distance between dissimilar functions. 
The training objective is to fine-tune the network so that the distance between the functions $F$ and the positive functions $P$ is closer than the distance between the functions $F$ and the negative functions $N$, which is illustrated below:

{\footnotesize
\begin{equation}
    max(||E_F - E_P|| - ||E_F - E_N|| + \epsilon, 0)
\end{equation}
}

where $E_F$,  $E_P$, and $E_N$ are the semantic embeddings of function $S$, $P$, and $N$ respectively. 
$\epsilon$ is the margin of the distance between $S$ and $N$. 
By default, $\epsilon$ is set to 1, which means the cosine distance between
{a function and its irrelevant function should be 1}.



\subsection{Vulnerability Explanation}
\label{lab:explainer}

Vulnerability explanation aims to provide developer-oriented natural language descriptions for problematic source code, which involves two aspects: building a code-related crowdsourced knowledge database and extracting key aspects for understanding vulnerable functions.

\subsubsection{Crowdsourced Knowledge Database}

This phase aims at managing diverse and useful information from developer forums (i.e., Stack Overflow) since the developer forums provide a lot of information in the form of question and answer (Q\&A) about (usually problematic) codes.
Meanwhile, users can also vote on the answers to distinguish the value of the questions and the corresponding answers.


In our knowledge database, we focus on two objects: questions/posts about a technical problem and answers for solving this problem.
For a question/post, it usually contains a title for concisely describing a problem, the details of the question, the source codes involved as well as an optional tag.
For an answer, it has a label to indicate whether it is a suggested one.
Meanwhile, the answer may give a detailed description about \textit{why it arises the problem, where the root cause exists, and how to solve it, especially for the suggested one}.
The descriptions are usually presented in the form of natural language, while the code presents the potential correctness solutions. 
The solution does not always work successfully for each user who is facing a similar problem because of environmental differences.
However, an explanation of problematic codes will inspire other users who encounter similar problems to understand the root cause.

Additionally, we connect posts with the same tags for retrieving answers more efficiently in the next phase (i.e., Results Explainer), as this process can fuse related posts with relevant problems.


\subsubsection{Result Explainer}

The crowdsourced code knowledge database helps to fuse useful information when addressing similar problems, while the result explainer aims at both retrieving relevant questions/posts involving similar source codes and extracting key aspects of the problems from the suggested answers.

The first step is to figure out the most (especially semantically)  relevant source codes explicitly.
In this paper, for retrieving the most semantically similar problematic functions, we adopt UniXcoder to obtain semantic embedding of functions since the model has been well pre-trained with contrastive learning technology.
Additionally, for a given retrieved post, there usually exists many responses from different users with varying experiences.
All the diverse responses can be useful since different developers may give their responses in different development environments (i.e., issues that occurred in Windows OS or Linux OS).
Therefore, apart from retrieving similar problematic functions, we also design an effective quality-first sorting strategy as follows to prioritize the most useful response/explanation.

{\footnotesize
\begin{equation}
\begin{split}
& Ranking\ Score =Func\_Sim \times (\frac{score_i}{\sum_{j=1}^{N}score_j} \times  Asps.)\\
&Asps. = 0.5\times I(Cau.) + 0.3\times I(Imp.) + 0.1\times I(Sol.) + 0.1\times I(Acpt.)
\end{split}
\end{equation}
}

where $Func\_Sim$ represents the similarity between the code in a given post and the vulnerable function, $score_i$ means the score of an answer $i$ in a post, which is voted by users.
A high score usually reflects the high quality of the answer.
$N$ represents the number of answers in the given post and $I(\cdot)$ is an indicator function.
It equals 1 if the condition is satisfied else it equals 0.
For example, $I(Cau.)$ equals 1 when the answer contains the root cause description to explain a problem.
In addition, it is possible that the root cause ($Cau.$), the impact ($Imp.$), and the potential solution ($Sol.$) provide different information for developers to understand the problems in codes.
Therefore, we assign different weights to indicate their priority.
Finally, if an answer is marked as \textit{Accept} ($Acpt.$), it means the answer has high quality for solving the problem, and we take it into consideration and assign the weight to 0.1.

The second step is to extract the key aspects for understanding the problem.
As introduced in the crowdsourced knowledge database, the suggested answer may contain a detailed description that explains key aspects (e.g., root cause, impact, solution, etc.) of the problem in source codes.
In our explanation model, we focus on the following three aspects: \textit{root cause}, \textit{impact}, and \textit{solution}, which are usually long clauses or sentences.

To extract the root cause, impact, and solution, we leverage the BERT-based Question Answering model~\cite{devlin2018bert,sun2021generating},  which is based on a pre-trained BERT model for retrieving questions and answers in a given content scope. 
The input of the model includes a question and the scope for answering the question.
The model outputs the start and end word index as the answer clause. 
In our application of BERT-QA, we adopt the $content$ of a suggested answer as the scope of question answering, and we input three what-is questions (i.e., ``\textit{what is root cause}'', ``\textit{what is impact}'' and ``\textit{what is the solution}'') into the model to find corresponding answers.
Benefiting from the language modeling capability of BERT, BERT-QA can handle complex clauses of the root cause, impact, and solution, and select the most appropriate information from the long response texts.

We train the BERT-QA model with 1,678 question-answer pairs (920 of reasons, 391 of impacts, and 492 of solutions), which is constructed manually from 55,627 posts (121,635 answers) in Stack Overflow. 
We build both positive and negative questions for which the answers can or cannot be found in the given posts. 
The negative questions help the model to learn when it fails to find any answer in the scope.
This characteristic is extremely important for extracting the root cause, impact, and solution since not all posts exactly and completely describe all three aspects. 
Otherwise, the BERT-QA model will have no ability to handle negative questions and extract some irrelevant content as the answer for a question.


\section{Experimental Design}
\label{sec:settings}

In this section, we first present features of the studied datasets, and then introduce the baseline approaches.
Following that, we describe the performance metrics as well as the experimental settings. 

\subsection{Datasets}
\label{lab:dataset}





\textbf{Vulnerability Dataset}.
We use the benchmark dataset provided by Fan et al.~\cite{fan2020ac} due to the following reasons. 
The first one is to establish a fair comparison with existing approaches (e.g., IVDetect, LineVul).
The second one is to evaluate whether existing approaches have a good generalization performance on detecting the fixed functions since Fan et al.~\cite{fan2020ac}'s dataset is the only one vulnerability dataset that provides the fixed version of vulnerable functions.
The last one is to satisfy the distinct characteristics of the real world as well as the diversity in the dataset, which is suggested by previous works~\cite{hin2022linevd,chakraborty2021deep}.

Fan et al.~\cite{fan2020ac} built the large-scale C/C++ vulnerability dataset named Big-Vul from Common Vulnerabilities and Exposures (CVE) database and open-source projects.
Big-Vul totally contains 3,754 code vulnerabilities collected from 348 open-source projects spanning 91 different vulnerability types from 2002 to 2019.
It has 188,636 C/C++ functions with a vulnerable ratio of 5.7\% (i.e., 10,900 vulnerability functions).
The authors linked the code changes with CVEs as well as their descriptive information to enable a deeper analysis of the vulnerabilities.
In our work, some baselines need to obtain the structure information (e.g., control flow graph (CFG), data flow graph (DFG)) of the studied functions.
Therefore, we adopt the same toolkit with \textit{Joern}~\cite{joern} to transform functions.
The functions are dropped out directly if they cannot be transformed by \textit{Joern} successfully.
We also remove the duplicated functions and the statistics of the studied dataset are shown in Table~\ref{tab:dataset}.

\begin{table}[htbp]
  \centering
  \caption{The statistic of studied dataset}
  \resizebox{\linewidth}{!}{
    \begin{tabular}{lrrrr}
    \toprule
    \textbf{Datasets} & \multicolumn{1}{l}{\textbf{\# Vul.}} & \multicolumn{1}{l}{\textbf{\# Non-Vul.}} & \multicolumn{1}{l}{\textbf{\# Total}} & \multicolumn{1}{l}{\textbf{\% Vul.: Non-Vul.}} \\
    \midrule
    Original Big-Vul & 10,900  & 177,736  & 188,636  & 0.061 \\
   \rowcolor{lightgray} Filtered Big-Vul  & 5,260  & 96,308  & 101,568  & 0.055 \\
    \midrule
    Training & 4,208  & 4,208  & 8,416  & 1 \\
    Validating & 526   & 9,631  & 10,157  & 0.055 \\
    Testing & 526   & 9,631  & 10,157  & 0.055 \\
    \bottomrule
    \end{tabular}%
    }
  \label{tab:dataset}%
\end{table}%

\textbf{Crowdsourced Dataset}.
Apart from the widely used vulnerability dataset, we also need to build a crowdsourced dataset manually in order to provide explanations for the detected vulnerabilities. 
In this paper, we crawl posts as well as their answers from Stack Overflow, where the posts are labeled with C or C++ and there is at least one code snippet in their content.
Finally, we obtain 55,627 posts with 121,635 answers, which are further used to build a knowledge database.

\subsection{Baselines}

To comprehensively compare the performance of \toolname~ with existing work, in this paper, we consider the four SOTA approaches: Devign~\cite{zhou2019devign}, {\sc ReVeal}~\cite{chakraborty2021deep}, {\sc IVDetect}~\cite{li2021vulnerability}, and LineVul~\cite{fu2022linevul}.
We briefly introduce them as follows.

\textbf{Devign} proposed by Zhou et al.~\cite{zhou2019devign} is a general graph neural network based model for graph-level classification through learning on a rich set of code semantic representations including AST, CFG, DFG, and code sequences.
It uses a novel $Conv$ module to efficiently extract useful features in the learned rich node representations for graph-level classification.
    
{{\sc \textbf{ReVeal}} proposed by Chakraborty et al.~\cite{chakraborty2021deep} contains two main phases: feature extraction and training. 
In the former phase, {\sc ReVeal} translates code into a graph embedding, and in the latter phase, {\sc ReVeal} trains a representation learner on the extracted features to obtain a model that can distinguish the vulnerable functions from non-vulnerable ones.}

{{\sc \textbf{IVDetect}} proposed by Li et  al.~\cite{li2021vulnerability} involves two components: coarse-grained vulnerability detection and fine-grained interpretation.
As for vulnerability detection, they process the vulnerable code and the surrounding contextual code in a function distinctively, which can help to discriminate the vulnerable code and the benign ones.
In particular, {\sc IVDetect} represents source code in the form of a program dependence graph (PDG) and treats the vulnerability detection problem as graph-based classification via graph convolution network with feature attention.
As for interpretation, {\sc IVDetect} adopts a GNNExplainer to provide fine-grained interpretations that include the sub-graph in PDG with crucial statements that are relevant to the detected vulnerability.}
    
\textbf{LineVul} proposed by Fu et al.~\cite{fu2022linevul} is a Transformer-based line-level vulnerability prediction approach. 
LineVul leverages BERT architecture with self-attention layers which can capture long-term dependencies within a long sequence. 
Besides, benefiting from the large-scale pre-trained model, LineVul can intrinsically capture more lexical and logical semantics for the given code input.
Moreover, LineVul adopts the attention mechanism of BERT architecture to locate the vulnerable lines for finer-grained detection.

\subsection{Evaluation Measures}

To evaluate the effectiveness of \toolname on vulnerability detection, we consider the following five metrics: Accuracy, Precision, Recall, F1-score, and PR-AUC.

\textbf{Accuracy} evaluates the performance that how many functions can be correctly labeled. It is calculated as:
$\frac{TP+ TN}{TP+FP+TN+FN}$.

\textbf{Precision} is the fraction of true vulnerabilities among the detected ones. It is defined as:
$\frac{TP}{TP+FP}$.

\textbf{Recall} measures how many vulnerabilities can be correctly detected. It is defined as: 
$\frac{TP}{TP+FN}$.

\textbf{F1-score} is a harmonic mean of $Precision$ and $Recall$ and can be calculated as:
    $\frac{2 \times P \times R}{P + R}$.

\textbf{PR-AUC} is the area under the precision-recall curve and is a useful metric of successful prediction when the class distribution is very imbalanced~\cite{hin2022linevd}. 
The precision-recall curve shows the trade-off between precision and recall for different thresholds. 
A high area under the curve indicates both high recall and high precision, where high precision corresponds to a low false positive rate, and high recall corresponds to a low false negative rate. 

\subsection{Experimental Setting}

We implement our vulnerability detection and explanation model \toolname in Python with the help of PyTorch framework.
Besides, we utilize \textit{unixcoder-base-nine}~\cite{guo2022unixcoder} from Huggingface~\cite{huggingface} as our basic model, which is a pre-trained model on NL-PL pairs of CodeSearchNet dataset and additional 1.5M NL-PL pairs of C, C++, and C\# programming language. 
We fine-tune \toolname on the studied datasets to obtain a set of suitable parameters for the vulnerability detection task and fine-tune BERT-QA model on the manually labeled question-answer datasets.
All the models are fine-tuned on four NVIDIA GeForce RTX 3090 graphic cards.
During the training phase, we use $\mathit{Adam}$ with a batch size of 32 to optimize the parameters of \toolname.
We also leverage $\mathit{GELU}$ as the activation function. 
A dropout of 0.1 is used for dense layers before calculating the final probability. 
We set the maximum number of epochs in our experiment as 20 and adopt an early stop mechanism to obtain good parameters.
The models (i.e., \toolname and baselines) with the best performance on the validation set are used for the evaluations.




\section{Experimental Results}
\label{sec:results}

To investigate the feasibility of \toolname~ on software vulnerability detection and detection result explanation, our experiments focus on the following four research questions:
\begin{itemize}[leftmargin=*]

\item \textbf{RQ-1}. {\em To what extent can the function-level vulnerability detection performance \toolname~ achieve?}

\item \textbf{RQ-2}. {\em How does the paired instance building strategy impact the performance of \toolname?}

\item \textbf{RQ-3}. {\em How does the size of paired instance impact the performance of \toolname?}

\item \textbf{RQ-4}. {\em How well does \toolname~ perform on explaining the detection results?}
\end{itemize}

In RQ1, we aim to investigate the performance of the \toolname on vulnerability detection by considering it with SOTA baselines (cf. Section~\ref{lab:rq1}).
In RQ2 and RQ3, we explore the impact of design options of contrastive learning on the performance of \toolname (cf. Section~\ref{rq2}, \ref{sec:rq3}).
In RQ4, we explore the \toolname's usefulness for helping developers understand vulnerable functions (cf. Section~\ref{sec:rq4}).

    

    



\subsection{\bf [RQ-1]: Effectiveness on Vulnerability Detection.}
\label{lab:rq1}

\noindent
\textbf{Objective.}
Benefiting from the powerful representation capability of deep neural networks, many DL-based vulnerability detection approaches have been proposed~\cite{zhou2019devign,li2021vulnerability}.
However, as vulnerable functions are usually fixed with a few modifications (52.6\% vulnerable functions can be fixed within 5 (76.7\% for 10) lines of codes), they have subtle lexical differences with the non-vulnerable functions.
Existing SOTA deep learning approaches (i.e., Devign, {\sc ReVeal}, {\sc IVDetect}, etc.) cannot perform well on the fixed functions (non-vulnerable).
The main reason falls into the limitations of effective semantic embedding among lexical similar functions.
In this paper, we propose a novel approach \toolname, which is built on a contrastive learning framework with a pre-trained model as a semantic encoder as suggested by previous work~\cite{chen2022varclr}.
The experiments are conducted to investigate whether \toolname~ outperforms SOTA function-level vulnerability detection approaches.


\noindent
\textbf{Experimental Design.}
We consider the four SOTA baselines:
Devign~\cite{zhou2019devign}, {\sc ReVeal}~\cite{chakraborty2021deep}, {\sc IVDetect}~\cite{li2021vulnerability}, and LineVul~\cite{fu2022linevul}.
These approaches can be divided into two categories: GNN-based one (i.e., Devign, {\sc ReVeal} and {\sc IVDetect}) and Pre-trained-based one (i.e., LineVul).
Besides, in order to comprehensively compare the performance among baselines and \toolname, we consider five widely used performance measures and conduct experiments on the popular dataset.
Since GNN-based approaches usually need to obtain the structure information of the function (e.g., CFG, DFG),  we adopt the same toolkit with \textit{Joern} to transform functions.
Finally, the filtered dataset (shown in Table~\ref{tab:dataset}) is used for evaluation.
We follow the same strategy to build the training data, validating data, and testing data from the original dataset with previous work does~\cite{fu2022linevul, ni2022defect}. 
Specifically, 80\% of functions are treated as training data, 10\% of functions are treated as validation data, and the left 10\% of functions are treated as testing data. 
We also keep the distribution as same as the original ones in training, validating, and testing data.

Meanwhile, for a specific function, \toolname needs to select appropriate positive instances and negative instances.
For the positive instances, we adopt the different embedding vectors of the same function by randomly dropping out some weights in the network of the semantic encoder.
For the negative instances, we consider all the other instances (i.e., functions) in the same mini-batch with the given instance and use the average semantic vector representation.
We consider three types of paired instances selection strategies (i.e., \textit{SimCL}, \textit{SimDFE} and \textit{R-Drop}. cf. Section~\ref{rq2}), and in this RQ, we adopt the \textit{R-Drop} strategy since it has overall best performance.

Finally, since our target is to build an effective vulnerability detection model, especially for discriminating lexically similar but semantically distinct functions, we further conduct an analysis on how \toolname performs on the fixed version of vulnerable functions in the testing dataset.

\begin{table}[htbp]
  \centering
  \caption{Vulnerability detection results of \toolname compared against four baselines.}
  \resizebox{\linewidth}{!}
  {
    \begin{tabular}{lrrrr}
    \toprule
    \textbf{Methods} & \textbf{F1-score} & \textbf{Recall} & \textbf{Precision} & \textbf{PR-AUC} \\
    \midrule
    Devign & 0.200  & \cellcolor{lightgray}\textbf{0.660} & 0.118  & 0.115  \\
    {\sc ReVeal} & 0.232  & 0.354  & 0.172  & 0.145  \\
    {\sc IVDetect} & 0.231  & 0.540  & 0.148  & 0.177  \\
    LineVul & 0.272  & 0.620  & 0.174  & 0.233  \\
    \midrule
    \cellcolor{lightgray}\textbf{\toolname} & \cellcolor{lightgray}\textbf{0.336} & 0.414  &\cellcolor{lightgray} \textbf{0.282} & \cellcolor{lightgray} \textbf{0.270} \\
    \midrule
    \textit{{Improv.}} & {\makecell[r]{23.5\%-68.0\%}} & -- & {\makecell[r]{ 62.1\%-139.0\%}} & {\makecell[r]{15.9\%-134.8\%}} \\
    \bottomrule
    \end{tabular}%
    }
  \label{tab:rq1-1}%
\end{table}%

\noindent
\textbf{Results.}
The evaluation results are reported in Table~\ref{tab:rq1-1} and the best performances are highlighted in bold. 
According to the results, we find that our approach \toolname~ outperforms all SOTA baseline methods on almost all performance measures except \textit{Recall}. 
In particular, \toolname obtains 0.336, 0.282, and 0.270 in terms of F1-score, Precision, and PR-AUC, which improves baselines by 23.5\%-68.0\%, 62.1\%-139.0\%, and 15.9\%-134.8\% in terms of F1-score, Precision, and PR-AUC, respectively.

In terms of \textit{Recall}, Devign performs the best (0.660) and LineVul performs similarly with Devign (0.620), which means that both the pre-trained model and the GNN-based model can achieve better performance of \textit{Recall}. 

\vspace{2mm}
\begin{table}[!htbp]
  \centering
  \caption{The effectiveness of \toolname compared against four baselines on fixed functions in testing dataset}
  \resizebox{\linewidth}{!}{
    \begin{tabular}{lrr|rr}
    \toprule
    \textbf{Methods} & \textbf{\# Correct} & \textbf{Accuracy} & \textbf{\# Improv.} & \textbf{\% Improv.} \\
    \midrule
    Devign & 194   & 36.9\% & 125   & 64.4\% \\
    {\sc ReVeal} & 297   & 56.5\% & 22    & 7.4\% \\
    {\sc IVDetect} & 209   & 39.7\% & 110   & 52.6\% \\
    LineVul & 202   & 38.4\% & 117   & 57.9\% \\
    \midrule
    \rowcolor{lightgray} \textbf{\toolname} &  \textbf{319} & \  \textbf{60.6\%} &     \textbf{22-125}  &  \textbf{7.4\% - 64.4\%} \\
    \bottomrule
    \end{tabular}%
    }
  \label{tab:rq1-2}%
\end{table}%

\begin{figure*}[!htbp]
    \centerline{
    \includegraphics[width=.8\linewidth]{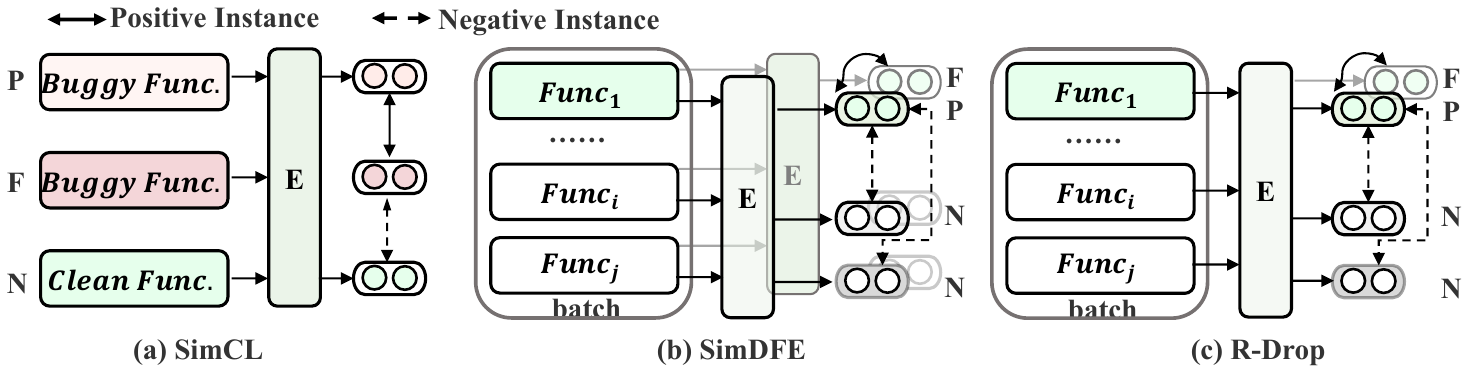}
    }
    \caption{Three different contrastive paired instances construction}
    \label{fig:paired-instances}
\end{figure*}

The performance comparisons of \toolname and four SOTAs on the fixed functions are presented in Table~\ref{tab:rq1-2}.
According to Table~\ref{tab:rq1-2}, we find that all SOTAs have poor performance on classifying the fixed versions (i.e., the clean version) in the testing dataset (i.e., 526 vulnerable functions), while \toolname~ can achieve the best performance.
More precisely, \toolname can correctly classify 319 fixed versions of functions as clean ones, which outperforms Devign (i.e., 194), {\sc ReVeal} (i.e., 297), {\sc IVDetect} (i.e., 209), and LineVul (i.e., 202) by 64.4\%, 7.4\%, 52.6\%, and 57.9\%, respectively.
The results indicate that \toolname has a better representation of learning ability than the four baselines.

\intuition{{\bf Answer to RQ-1}:
\toolname outperforms the SOTA baselines at the function-level software vulnerability detection.
Particularly, it achieves overwhelming results at both F1-score and PR-AUC, which indicates that \toolname equipped with contrastive learning as well as pre-trained model has a stronger ability to learn the semantics of functions, especially for those functions with lexical similarity but have distinct semantics.}

\subsection{\bf [RQ-2]: Impacts of Contrastive Paired Instances Construction.}
\label{rq2}

\noindent
\textbf{Objective.}
The contrastive learning framework needs to build triplet paired instances, which are used to measure how close the two similar instances are and how far the two dissimilar instances are.
Therefore, it is important to conduct a study on how the constructed positive instances and negative instances of a given function affect the learning of semantic representation.

\noindent
\textbf{Experimental Design.}
We consider three types (i.e., \textit{SimCL}, \textit{SimDFE}, and \textit{R-Drop}) of paired instances (i.e., positive instances and negative instances) building strategies to train our proposed approach \toolname.
The differences among these strategies are illustrated in Fig.~\ref{fig:paired-instances} and we introduce them in detail as follows.

\textbf{$\bullet$ SimCL} (simple contrastive learning) means building the negative instance of a vulnerable function with its corresponding fixed version. 
For its positive equivalent function, we input the original function twice into the same encoder with different weights (i.e.,  dropout used as noise) inside the model and obtain two embedded vectors.
The two vectors are interchangeably treated as positive instances.

\textbf{$\bullet$ SimDFE} (simple duplicate function embedding) means inputting all functions (noted as $f_1, f_2,\cdots, f_n$, $n$ is the size of batch) in a batch twice into the same encoder with different weights, which is inspired by~\cite{gao2021simcse}.
That is, each function will have two embedded vectors, noted as $f_{11}, f_{12}, f_{21}, f_{22}, \cdots, f_{n1}, f_{n2}$.
Take $f_{1}$ as an example, $f_{11}$ and $f_{12}$ are interchangeably treated as positive instances and $f_{ij}$ are treated as negative instances, where $i \in [2,n]$ and $j \in [1,2]$.
We use the average difference between $f_1$ and all negative instances as their dissimilarity.

\textbf{$\bullet$ R-Drop} (random dropout) means to input one function (noted as $f_1, f_2,\cdots, f_n$, $n$ is the size of batch) in a batch twice and the rest function in the same batch once into the same encoder.
For the given function embedded with an encoder twice, we adopt the random \textit{dropout} operation to the network to obtain the equivalent positive embedding.    
Take $f_{1}$ as an example, $f_{11}$ and $f_{12}$ are interchangeably treated as positive instances and $f_{i} (i\in [2,n])$ are treated as negative instances.
We use the average difference between $f_1$ with all negative instances  as their dissimilarity.

The experimental dataset is set the same as the experiment of RQ-1 (i.e., 80\% for training, 10\% for validating, and 10\% for testing). 
We also consider the five performance measures (i.e., Precision, Recall, F1-score, PR-AUC, and Accuracy) for comprehensively studying the impact of different paired instances building strategies. 
Additionally, in this study, we set the batch size $n$ as 32.

\begin{table}[htbp]
  \centering
  \caption{The performance difference among three different paired instances construction strategies}
  \resizebox{\linewidth}{!}{
    \begin{tabular}{l|cccc|cc}
    \toprule
    \multirow{2}[4]{*}{\textbf{Strategy}} & \multicolumn{4}{c|}{\textbf{Testing Data}} & \multicolumn{2}{c}{\textbf{Fixed function}} \\
\cmidrule{2-7}          & \textbf{F1-score} & \textbf{Recall} & \textbf{Precision} & \textbf{PR-AUC} & \textbf{\# Num} & \textbf{Accuracy} \\
    \midrule
    \rowcolor[rgb]{ .949,  .949,  .949} \textit{\toolname} & \textit{0.303} & \textit{\textbf {0.536}} & \textit{0.211} & \textit{0.245} & \textit{243} & \textit{0.462} \\
    \textbf{\toolname$_{SimCL}$} & 0.313  & 0.504  & 0.227  & 0.257  & 269   & 0.511  \\
    \textbf{\toolname$_{SimDFE}$} & 0.324  & 0.481  & 0.244  & 0.265  & 268   & 0.510  \\
   \cellcolor{lightgray}\textbf{\toolname$_{R-Drop}$} &\cellcolor{lightgray}\textbf{0.336}  & 0.414  & \cellcolor{lightgray}\textbf{0.282}  & \cellcolor{lightgray}\textbf{0.270} & \cellcolor{lightgray}\textbf{319}   & \cellcolor{lightgray}\textbf{0.606}  \\
    \midrule
    \textbf{\textit{Improv.}} & \makecell[c]{3.3\%\\to\\10.9\%} & ---     & \makecell[c]{7.6\%\\to\\33.6\%} & \makecell[c]{4.9\%\\to\\10.2\%} & \multicolumn{2}{c}{\makecell[c]{10.3\%\\to\\31.3\%}} \\
    \bottomrule
    \end{tabular}%
    }
  \label{tab:rq2}%
\end{table}%

\vspace{2mm}
\begin{figure}[htbp]
\centering
\includegraphics[width=\linewidth]{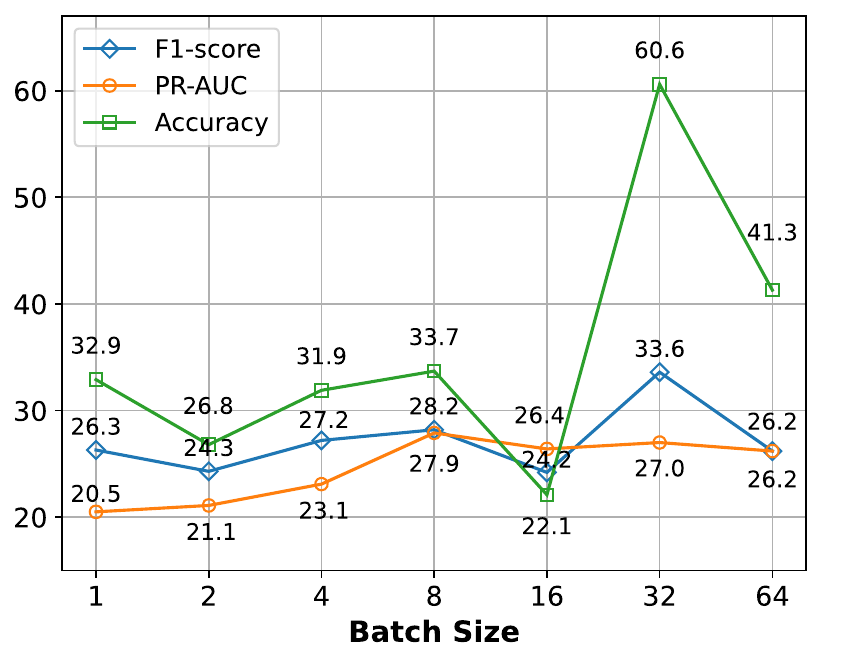}
\caption{The varying performance of \toolname with different batch size}
\label{fig:rq3}
\end{figure}

\noindent
\textbf{Results.}
The comparison results are reported in Table~\ref{tab:rq2} and the best performances are highlighted in bold for each performance measure. 
According to the results, we can obtain the following observations: 
(1) All paired instance construction strategies have the advantage of learning function semantic embedding in the scenario of vulnerability detection. 
Particularly, \textit{SimCL}, \textit{SimDFE}, and \textit{R-Drop} improve the baseline (UniXcoder without contrastive learning) by 3.3\%-10.9\%, 7.6\%-33.6\%, 4.9\%-10.2\%, and 10.3\%-31.3\% in terms of F1-score, Precision, PR-AUC, and Accuracy.
(2) The \textit{SimDFE} performs better than the \textit{SimCL} and the \textit{R-Drop} is the dominated one among the three strategies.
(3) The \textit{SimCL} performs worse than the other two strategies and the main reason may come from the small size of negative instances (i.e., only one negative instance), which limits the information for \toolname to discriminate the difference between positive instances and negative instances.
(4) The contrastive learning strategy, to some degree, can decrease the performance of \textit{Recall}.
However, it has an improvement on two comprehensive performance measures (i.e., \textit{F1-score} and \textit{PR-AUC}), especially for distinguishing two lexically similar functions with distinct semantics (i.e., an improvement on \textit{Accuracy}).

\intuition{{\bf Answer to RQ-2}: All paired instance construction strategies present their own advantages in learning function semantic embedding and the \textit{R-Drop} strategy performs the best.
}

\vspace{2mm}
\begin{figure*}[htbp]
\centering
\subfigure[Devign]{
\begin{minipage}[t]{0.16\linewidth}
\centering
\includegraphics[width=1in]{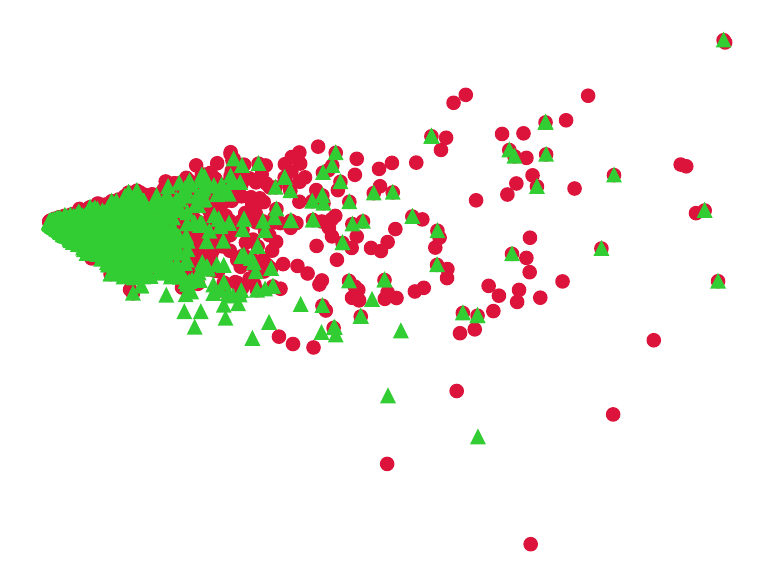}
\end{minipage}%
}%
\subfigure[{\sc ReVeal}]{
\begin{minipage}[t]{0.16\linewidth}
\centering
\includegraphics[width=1in]{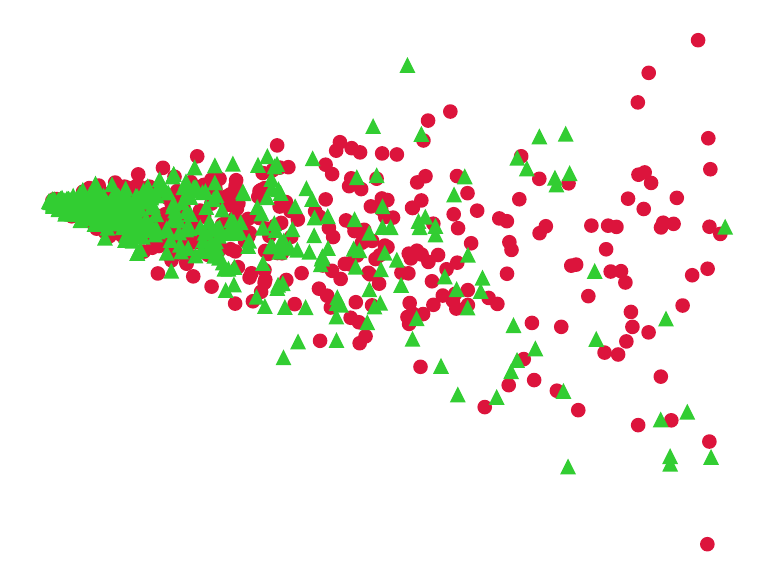}
\end{minipage}%
}%
\subfigure[{\sc IVDetect}]{
\begin{minipage}[t]{0.16\linewidth}
\centering
\includegraphics[width=1in]{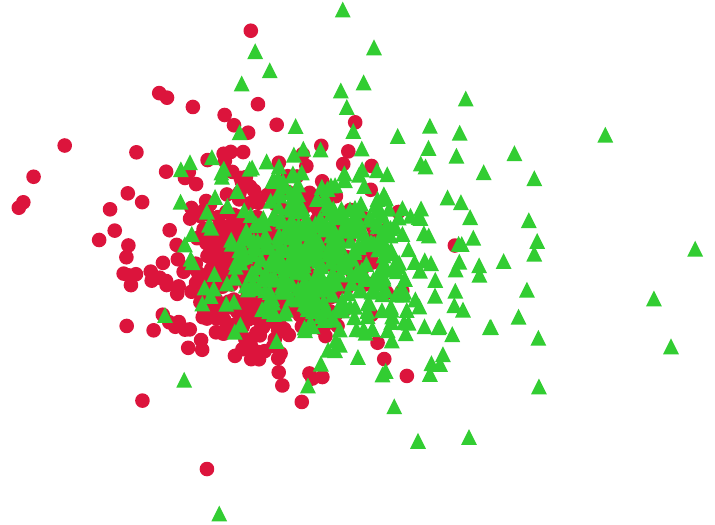}
\end{minipage}
}%
\subfigure[LineVul]{
\begin{minipage}[t]{0.16\linewidth}
\centering
\includegraphics[width=1in]{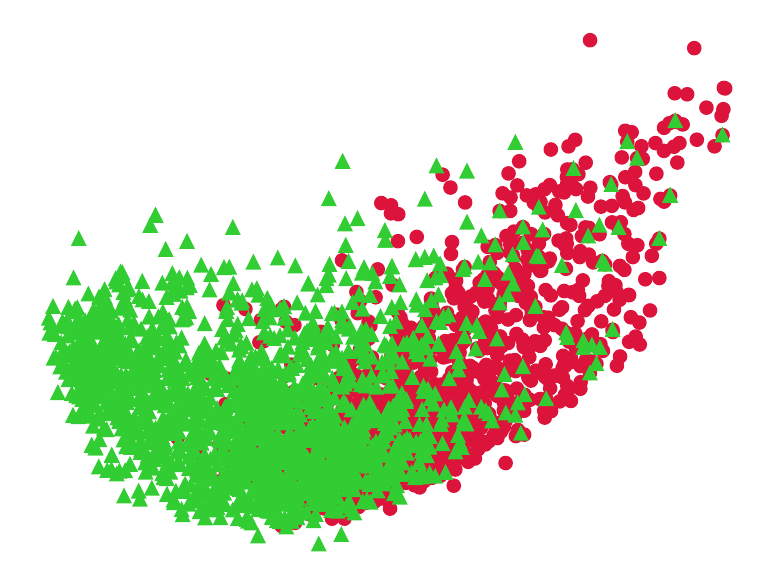}
\end{minipage}
}%
\subfigure[UniXcoder]{
\begin{minipage}[t]{0.16\linewidth}
\centering
\includegraphics[width=1in]{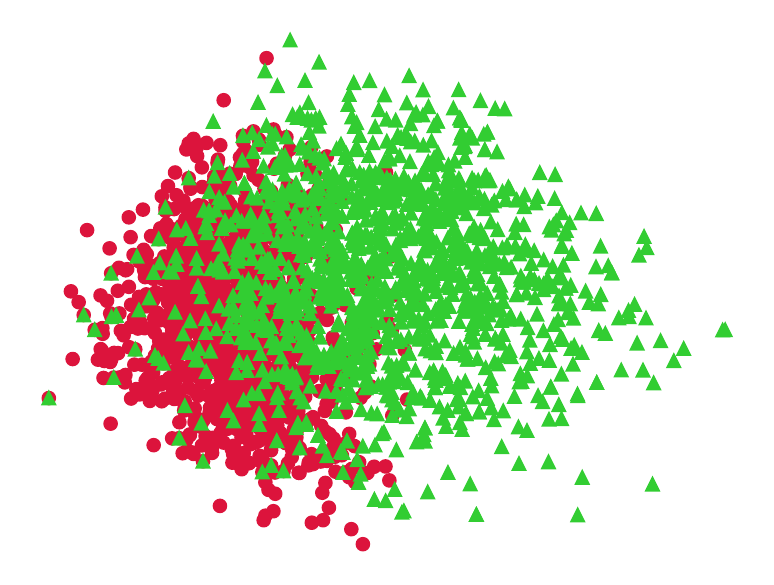}
\end{minipage}
}%
\subfigure[\toolname]{
\begin{minipage}[t]{0.16\linewidth}
\centering 
\includegraphics[width=1in]{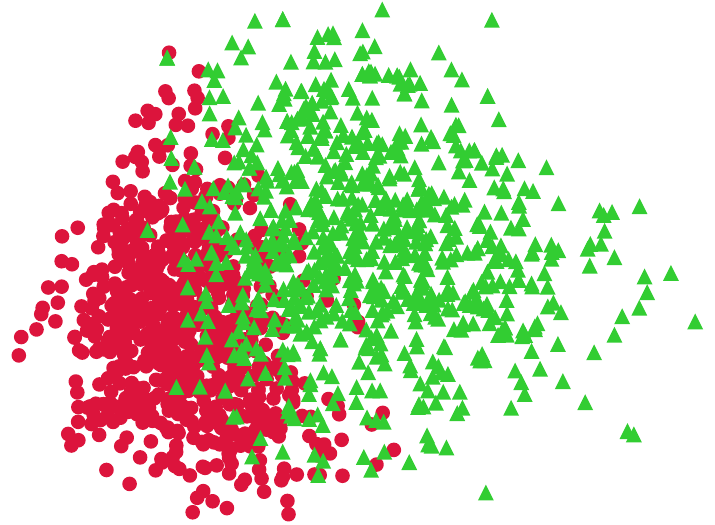}
\end{minipage}
}%
\centering
\caption{Visualization of the separation between vulnerable (denoted by \textcolor{red}{\ding{108}}) and non-vulnerable (denoted by \textcolor{green}{\ding{115}}).}
\label{fig:disscusion1}
\end{figure*}

\subsection{\bf [RQ-3]: Impacts of Paired Instances Size.}
\label{sec:rq3}

\noindent
\textbf{Objective.}
In RQ-2, we find that the number of negative instances has an impact on \toolname's performance of learning semantic embedding.
Therefore, we want to conduct a deeper experiment on how the batch size (i.e., the number of negative instances) impacts the performance of \toolname~ on discriminating dissimilar instances from similar ones.

\noindent
\textbf{Experimental Design.}
According to RQ-2, we find that the \textit{R-Drop} strategy has an overall better performance than the others.
Meanwhile, considering the fact that the larger the batch size is, the more memory \toolname consumes, we re-run \toolname with \textit{R-Drop} strategy on the following varying settings of batch size: 1, 2, 4, 8, 16, 32, and 64.
Because of the limitation of graph memory (i.e., four NVIDIA RTX 3090) and the size of functions, we cannot perform larger batch sizes (i.e., 128 or 256). 
Besides, the experimental dataset is set as same as that in previous RQs.
We evaluate the performance of \toolname on testing data with two comprehensive performance measures (i.e., F1-score and PR-AUC), and we adopt Accuracy to evaluate the performance on the fixed version of vulnerable functions.

\noindent
\textbf{Results.}
The evaluation results of \toolname with varying batch size are illustrated in Fig.~\ref{fig:rq3}. 
According to the results, we have the following research findings:
(1) Different number of negative instance has varying impact on \toolname's performance.
(2) Almost all the metrics of \toolname (except Accuracy) go up with the increasing of negative instances when batch size is no larger than 32.
When batch size equals 64, all the performances drop to different degrees.
(3) Larger batch size may not lead to better performance and assigning a batch size of 32 is a good choice.

\intuition{{\bf Answer to RQ-3}: 
The number of negative instances has an impact on \toolname's performance and the larger number may not always guarantee better performance.
In our setting, a median size (i.e., 32) is more appropriate.}

\subsection{\bf [RQ-4]: Usefulness for Developers.}
\label{sec:rq4}

\noindent
\textbf{Objective.}
Though many novel approaches have been proposed and indeed achieved remarkable performance, existing methods cannot provide a developer-oriented, natural language-described explanation.
For example, \textit{what is the possible root cause of such vulnerability?}
Such types of explanations may (at least intuitively) help developers understand the identified vulnerability better.
However, considering the concealment of software vulnerabilities, we cannot observe two identical vulnerabilities. 
It is possible that similar/homogeneous vulnerabilities have similar root causes or lead to similar impacts.
Meanwhile, many publicly available developer forums (i.e., Stack Overflow) share similar problems and their responses may provide understandable natural language explanations about the issues. 
Therefore, we want to further utilize this useful and diverse information to provide participants with detailed explanations about the identified problematic codes.

\noindent
\textbf{Experimental Design.}
We first crawl posts labeled with C/C++ from Stack Overflow and build a database (cf. Section~\ref{lab:dataset}) to fuse all crowdsourced knowledge for retrieving important explainable information.
Considering that our work focuses on code-related problems, we filter those posts with no code snippet in their post content since the code snippet is the critical connective element when retrieving similar problematic codes.
In addition, for retrieving the most semantically similar problematic functions, we adopt \toolname to obtain semantic embedding of both vulnerable function and code snippet in post since our model has been well pre-trained with contrastive learning technology.
Then, we adopt the designed quality-first sorting strategy (cf. Section~\ref{lab:explainer}) to prioritize the retrieved answers.
Finally, the well pre-trained BERT-QA model (cf. Section~\ref{lab:explainer}) is adopted to extract three optional important descriptions (i.e., \textit{root cause}, \textit{impact}, and \textit{solution}) inside the answer.

Finally, we randomly select 20 vulnerable functions in testing datasets and invite 10 developers from a prominent IT company who have 5 to 8 years of experience in software security as our participants.
Each developer is asked to finish an experiment task that includes two vulnerable functions as well as their corresponding explanation recommended by \toolname. 
We evaluate the usefulness of our approach by analyzing the answers to the following questions given by participants.
More precisely, \toolname presents each vulnerable function with five retrieved answers from crowdsourced knowledge.
\begin{itemize}[leftmargin=*]
    \item Q1: Is the explanation related to the vulnerable function?
    \item Q2: Is the explanation comprehensive (i.e., the root cause, the impacts, and the suggestion. score: 1-5', 1(low)-3(middle)-5(high))? Which part is most important?
    \item Q3: Is the explanation useful to understand the vulnerability?
    \item Q4: In which result do you find the most desired answer? (score: 0-5', 0 means no desired answer.)
    \item Q5: Please sort the explanations according to their usefulness.
\end{itemize}
For Q1 and Q2, we aim to verify the relatedness and comprehensiveness of \toolname's recommendation.
For Q3 and Q4, we aim to evaluate the usefulness of \toolname, and Q5 is designed to evaluate the difference between the recommendations and developers' expectations.

\noindent
\textbf{Results.}
In Q1, except for 5 negative responses (i.e., providing unrelated explanations), 15 responses are positive to indicate the relatedness of recommended posts.
In Q2, the majority (i.e.,19/20 with larger than 3') agrees that \toolname's recommended posts provide the reasons (root cause) for problematic codes.
Besides, all responses (i.e., $\geq3$) are positive with the \textit{suggestion}.
However, about half of the participants give less than 3 scores to the \textit{impacts} of problematic code, which is consistent with our manually labeled data (the impacts of problematic code have the least number).
Meanwhile, everyone believes that giving the explanation of \textit{root cause} is most important for explaining a problematic code.
In Q3, 13 participants agree that the explanation extracted by \toolname can help them intuitively understand the vulnerable code and the remaining 7 responses have negative feedback, which also confirms the concealment of vulnerability.
In Q4, we find that 13 responses rank at top-3 (5 for top-1, 5 for top-2, and 3 for top-3), and 3 responses are scored with 0, which means that none of the recommended explanations are related to the vulnerable function.
Finally, in Q5, we use Mean Average Precision (MAP)~\cite{li2021vulnerability} to qualify the gap between our recommendation and developers' expectations. 
We get 0.565 of MAP, which means \toolname, to some degree, can give an acceptable recommendation list.

We analyze the negative responses about \toolname and find that the biggest problem falls into the completeness of our dataset, as \toolname cannot find the most semantic similar problematic codes with vulnerable functions in the built dataset (i.e., similarity $<0.5$).

\intuition{{\bf Answer to RQ-4}: Our user study reveals, to some extent, that \toolname presents the potential feasibility of assisting developers to intuitively understand the detected vulnerability. }











\section{Discussion}
\label{sec:discussion}

This section discusses open questions regarding the performance and threads to the validity of \toolname.

\subsection{Why \toolname outperforms Existing Baselines?}

DL-based vulnerability detection approaches have a strong ability to learn a feature representation to distinguish vulnerable functions and non-vulnerable ones.
Therefore, the efficacy of the models' vulnerability detection depends largely on how separable the feature representation of the two types of functions (i.e., vulnerable and non-vulnerable) are. 
The greater the separability of the two functions, the easier it is for a model to distinguish between them.


We adopt Principal Components Analysis (PCA)~\cite{abdi2010principal}  to inspect the separability of the studied models. 
PCA is a popular dimensionality reduction technique and is suited for projecting the original feature embedding into two principal dimensional embeddings.
Besides, we randomly sample the same number of non-vulnerable functions with vulnerable functions in the testing dataset for more clear visualization.

Fig.~\ref{fig:disscusion1} illustrates the separability of the studied approaches. 
From the visualization results (Fig.~\ref{fig:disscusion1}(a)–(d)), we can see that the majority of the functions are mixed and the boundary of each function is not clear, which indicates the difficulty of baselines in drawing the decision boundary.
In contrast, UniXcoder (shown in Fig.~\ref{fig:disscusion1}(e)) has better separability than baselines, which indicates the large-scale pre-trained language model (specially trained on C/C++ codes) has a stronger ability to understand the semantic of codes.
Lastly, Fig.~\ref{fig:disscusion1}(f) shows the separability of our \toolname. 
We can observe that \toolname has the best performance in distinguishing vulnerable functions from non-vulnerable ones.
Equipped with contrastive learning, \toolname can learn better semantic embedding of functions.

\subsection{Threats to Validity}

\noindent
{\textbf{Threats to Internal Validity} 
mainly correspond to  the potential mistakes in the implementation of our approach and other baselines. 
To minimize such a threat, we first implement our model by pair programming and directly utilize the pre-trained models for building vulnerability detectors.
We also use the original source code of baselines from the GitHub repositories shared by corresponding authors and use the same hyperparameters in the original papers. 
The authors also carefully review the experimental scripts to ensure their correctness.
}

\noindent
{\textbf{Threats to External Validity} mainly correspond to the studied dataset.
Even though we have evaluated models on those widely used vulnerability datasets in literature to ensure a fair comparison with baselines, the diversity of projects is also limited in the following aspects.
Firstly, all the studied projects (i.e., functions) are developed in C/C++ programming language. 
Therefore, projects developed in other popular programming languages (e.g., Java and Python) have not been considered.
Secondly, all the studied datasets are collected from open-source projects, and the performance of \toolname on commercial projects is unknown.
Thus, more diverse datasets should be collected and explored in future work.
}

\noindent
\textbf{Threats to Construct Validity} mainly correspond to  the performance metrics used in our evaluations.
To minimize such a threat, we adopt a few performance metrics widely used in existing work.
In particular, we totally consider five performance metrics including Accuracy, Precision, Recall, F1-score, and PR-AUC.

\section{Related Work}
\label{sec:relatedwork}
\subsection{AI-based Software Vulnerability Detection}

Software vulnerability detection has attracted much attention from researchers and many DL-based approaches have been proposed to automatically learn the vulnerability patterns from historical data~\cite{yamaguchi2014modeling,li2018vuldeepecker,zhou2019devign,li2021vuldeelocator,duan2019vulsniper,lin2017poster,chakraborty2021deep}, since the powerful learning ability of deep neural networks has been verified in many software engineering scenarios~\cite{ni2022best,zhu2019automatic,monperrus2018automatic} (e.g., defect prediction, defect repair).

Dam et al.~\cite{dam2017automatic} proposed a vulnerability detector with LSTM-based architecture. 
Russell et al.~\cite{russell2018automated} proposed another RNN-based architecture to automatically extract features from source code for vulnerability detection. 
However, these approaches assume source code is a sequence of tokens, which ignores the graph structure of the source code.
Therefore, Li et al.~\cite{li2021sysevr,li2018vuldeepecker} sequentially proposed two slice-based vulnerability detection approaches, VulDeePecker~\cite{li2018vuldeepecker} and SySeVR~\cite{li2021sysevr}, to learn the syntax and semantic information of vulnerable code. 
Following that, many graph neural network (GNN) based models~\cite{yamaguchi2014modeling,zhou2019devign} are proposed.
Cheng et al.~\cite{cheng2021deepwukong} proposed DeepWukong by embedding both textual and structural information of code into a comprehensive code representation.
Wang et al.~\cite{wang2020combining} proposed FUNDED by combining nine mainstream graphs. 
Cao et al.~\cite{cao2022mvd} proposed MVD to detect fine-grained memory-related vulnerability.

Apart from the coarse-grained models (e.g., function level), researchers also proposed many fine-grained models. 
Li et al.~\cite{li2021vuldeelocator} proposed VulDeeLocator by adopting a program slicing technique to narrow down the scope of vulnerability-prone lines of code. 
Fu et al.~\cite{fu2022linevul} proposed LineVul by leveraging the attention mechanism inside the BERT architecture for line-level vulnerability detection.
Hin et al.~\cite{hin2022linevd} proposed LineVD to formulate statement-level vulnerability detection as a node classification task. 

Different from previous work, our paper focuses on the effective semantic embedding of functions, especially those that are lexically similar.

\subsection{Interpretation for AI-based Software Vulnerability Detection}



Developing explainable models is one of the ways for vulnerability detection, which could provide a fine-grained vulnerability prediction outcome. 
Specifically, many works have attempted to detect line-level information by leveraging explainable AI for software engineering tasks, such as detecting source code lines for defect prediction~\cite{pornprasit2022deeplinedp,ni2022best}.
This raises the importance of research for interpretable AI-based models.

However, existing studies are limited to providing partial information for the explanation generation.
Zou et al.~\cite{zou2021interpreting} introduced a high-fidelity token-level explanation framework, which aims at identifying a small number of tokens that make significant contributions to a detector's prediction.
Li et al.~\cite{li2021vuldeelocator} proposed VulDeeLocator to simultaneously achieve high detection capability and high locating precision and it explains detection results at intermediate code.
Ding et al.~\cite{ding2021velvet} proposed a statement-level model via localizing the specific vulnerable statements with the assumption of receiving vulnerable source codes at the function level.  
Li et al.~\cite{li2021vulnerability} adopted explainable GNN to propose IVDetect and provided fine-grained interpretations.
Fu et al.~\cite{fu2022linevul} proposed a transformer-based line-level model named  LineVul and leveraged the attention mechanism of BERT architecture to explain the vulnerable code lines.
Recently, Sun et al.~\cite{sun2023slient} conducted the first research work on the application of Explainable AI in silent dependency alert prediction, which opens the door to the related domains.

Different from existing works that focus on explaining why AI-models give out the predicted results, our paper aims at making an explanation for the detected results by providing a develop-oriented natural language described explanation in order to heuristically help developers understand the root cause of the detected vulnerabilities.

\section{Conclusion and Future Work}
\label{sec:conclusion}

This paper proposes a novel approach \toolname, which is a function-level subtle semantic embedding for vulnerability detection along with heuristic explanations, technically based on pre-trained semantic embedding as well as contrastive learning.
\toolname firstly adopts contrastive learning to train the UniXcoder semantic embedding model for learning distinguishing semantic representation of functions regardless of their lexically similar information.
\toolname secondly builds a knowledge-based crowdsource dataset by crawling problematic codes in Stack Overflow to provide developers with heuristic explanations of the detected problematic codes.
The experimental results show the effectiveness of \toolname by comparing it with four SOTA deep learning-based approaches.

Our future work will investigate the generalization of contrastive learning to existing deep learning approaches for vulnerability detection.

\section{Data Availability}
The replication of this paper is publicly available~\cite{replication}.

\section*{Acknowledgements}{
This research is supported by the National Natural Science Foundation of China (No. 62202419), the Fundamental Research Funds for the Central Universities (No. 226-2022-00064), the Ningbo Natural Science Foundation (No. 2022J184), and the State Street Zhejiang University Technology Center.
}

\balance
\bibliographystyle{ACM-Reference-Format}
\bibliography{main}

\end{document}